\documentclass{aastex701}

\usepackage[T1]{fontenc}

\DeclareRobustCommand{\VAN}[3]{#2}
\let\VANthebibliography\thebibliography
\def\thebibliography{\DeclareRobustCommand{\VAN}[3]{##3}\VANthebibliography}

\usepackage{graphicx}	
\usepackage{amsmath}	
\usepackage{amssymb}	
\usepackage{amsfonts}
\usepackage[arrowdel]{physics}
\usepackage{graphicx}
\usepackage{float}
\usepackage{color, graphics}
\usepackage{verbatim}   
\usepackage{subfigure}  
\usepackage{acronym}
\usepackage{mathrsfs}
\usepackage{multirow}
\usepackage{cancel}
\usepackage{url}
\usepackage{slashed,afterpage}
\usepackage{units}
\usepackage{enumerate}  
\usepackage{mathtools}
\hypersetup{
    colorlinks=true,
    linkcolor=blue,
    filecolor=magenta,      
    urlcolor=cyan,
}
\def\beq{\begin{equation}\begin{aligned}}
\def\eeq{\end{aligned}\end{equation}}
\usepackage{tcolorbox}
\usepackage{graphicx}
\usepackage[english]{babel}
\usepackage{newtxtext,newtxmath}

\begin{document}

\title{Hint of dark matter-dark energy interaction in DESI DR2 and current cosmological dataset?}

\author[orcid=0000-0001-9716-7875,sname='Chakraborty']{Amlan Chakraborty}
\affiliation{Indian Institute of Astrophysics, Bengaluru, Karnataka 560034, India}
\affiliation{Department of Physics, Pondicherry University, R.V. Nagar, Kalapet, 605 014, Puducherry, India}
\email[show]{amlan.chakraborty@iiap.res.in}  

\author{Tulip Ray}
\affiliation{Indian Institute of Astrophysics, Bengaluru, Karnataka 560034, India}
\email{tulipray2804@gmail.com}

\author{Subinoy Das}
\affiliation{Indian Institute of Astrophysics, Bengaluru, Karnataka 560034, India}
\affiliation{Department of Physics, Pondicherry University, R.V. Nagar, Kalapet, 605 014, Puducherry, India}
\email[show]{subinoy@iiap.res.in}

\author{Arka Banerjee}
\affiliation{Department of Physics, Indian Institute of Science Education and Research, Homi Bhabha Road, Pashan, Pune 411008, India}
\email[show]{arka@iiserpune.ac.in}

\author{Vidhya Ganesan}
\affiliation{Indian Institute of Astrophysics, Bengaluru, Karnataka 560034, India}
\email{vidhya.gnsn@gmail.com}


\begin{abstract}

We present new constraints on an interacting dark matter–dark energy scenario motivated by string compactification, where a scalar field adiabatically tracks the minimum of an effective potential sourced by dark matter density. In this study, we focus on the Chameleon dark energy model and numerically solve the Klein-Gordon equation using a shooting algorithm to determine precise initial conditions such that the field rests at effective potential minima today. We perform a comprehensive Markov Chain Monte Carlo (MCMC) analysis using a combination of datasets, including Planck, BAO (SDSS and DESI DR2), Pantheon+, and SH$_0$ES. Our analysis shows a mild preference for a higher non-zero dark sector coupling, compared to earlier works on similar models, for two particular combinations of datasets: (i) Planck + DESI DR2 BAO +.Pantheon+, (ii) Planck + SDSS BAO + Pantheon+ + SH$0$ES. Notably, the inclusion of DESI DR2 and SH$0$ES data increases the inferred interaction strength to $\beta \sim 0.3$ (68\% C.L.) and yields weak and positive evidence in favor of the model over $\Lambda$CDM, with $\Delta\chi^2_{\rm min} = -4.75, -6.41$ and $\Delta$AIC= $-0.75, -2.41$ respectively. This model remains consistent with a phantom crossing at redshift $z\sim 0.5$, in agreement with the trend indicated by DESI observations. However, due to the settlement of the scalar field at the minima of the effective potential at the present epoch, the effective dark energy equation of state asymptotically approaches $w_{\rm eff}\to -1$. leading to only weak evidence in favor of this model when analyzed using the DESI DR2 dataset.

\end{abstract}



\section{\label{sec:level1}Introduction} 

The standard six-parameter phenomenological $\Lambda$CDM model has proven remarkably successful in accounting for the observed large-scale evolution of the universe \cite{2020,2022MNRAS.512.5657S,Sevilla_Noarbe_2021,PhysRevD.108.123519,PhysRevD.108.123518}. Nonetheless, with the advent of precision cosmology, several persistent discrepancies have emerged between different observational datasets. Notably, the so-called ``Hubble tension" and the``$S_8$ tension" have raised questions about the completeness of the $\Lambda$CDM framework \cite{2017NatAs...1E.121F,10.1093/mnras/stt601}. From a theoretical perspective, the exceedingly small value of the cosmological constant ($\Lambda$) continues to pose a profound challenge for fundamental physics \cite{RevModPhys.61.1}. Although current data remain broadly consistent with a cosmological constant as the source of dark energy, cosmological observations have yet to shed light on the physical origin or naturalness of its tiny value.

As an alternative to the cosmological constant, the possibility of a dynamical dark energy component, characterized by a time-dependent equation of state, has been extensively explored \cite{PhysRevLett.80.1582,PhysRevD.59.023509,Sharma_2022}. Such scenarios naturally lead to distinctive imprints on the growth of large-scale structures, offering potential observational signatures that deviate from the predictions of the standard $\Lambda$CDM model. These models are increasingly within the reach of present and forthcoming cosmological surveys, which are capable of probing the expansion history and structure formation with unprecedented precision. Notably, recent results from the Dark Energy Spectroscopic Instrument (DESI) have indicated mild tensions with the $\Lambda$CDM baseline, hinting at the possibility of a departure from a pure cosmological constant behavior \cite{DESI:2024mwx,2025desidr2,lodha2025, RoyChoudhury:2024wri,RoyChoudhury:2025dhe, RoyChoudhury:2025iis}. This, in turn, strengthens the case for exploring more general classes of dark energy models.
A particularly intriguing avenue in this context involves the presence of interactions within the dark sector—specifically, between dark energy and dark matter \cite{WETTERICH1988668,1995A&A...301..321W}.In the absence of any fundamental symmetry forbidding such couplings, interactions between a scalar field dark energy and dark matter fermions are not only theoretically allowed but may indeed arise naturally in particle physics-inspired frameworks \cite{WETTERICH1988668}. Initially, these type of model were introduced to address the coincidence problem \cite{PhysRevD.62.043511} and extensively investigated to get a statistically significant detection of the interaction in the dark sector; however, only upper bounds were found \cite{Amendola:2002bs, Amendola:2003eq, Amendola:2000ub}. These interactions can leave observable footprints on a wide range of cosmological probes, including the cosmic microwave background, large-scale structure, and expansion history that can be probed with recent datasets with higher precision. Consequently, numerous scalar-tensor theories have been proposed in the literature to model such interacting scenarios, opening up a rich phenomenological landscape to explore \cite{PhysRevLett.64.123,Casas_1992,Anderson_1998,PhysRevD.64.043509,PhysRevD.64.123516,PhysRevLett.89.081601,Comelli_2003,Amendola_2003,PhysRevD.67.083513,PhysRevD.68.023514,PhysRevD.69.063517,PhysRevLett.93.171104,PhysRevD.69.044026,PhysRevD.70.123518,Farrar_2004,Gubser_2004,PhysRevLett.93.091801,Olivares_2005,PhysRevD.71.043504,Peccei_2005,Fardon_2006,von_Marttens_2020,Baker:2022mby,Zhang:2022mdb,Piedipalumbo:2023dzg,Brax:2023qyp,Kading:2023mdk,Singh:2023zbm,Burgess_2022,Brax:2022vlf, van_der_Westhuizen_2024,DIVALENTINO2020100666,Bolotin:2013jpa, PhysRevD.101.123506, PhysRevD.102.123502, PhysRevD.101.063502, 10.1093/mnrasl/slaa207, 10.1093/mnras/stac229, LI2020135141, Bachega_2020,Li_2020, Mukhopadhyay:2019jla, vonMarttens:2019wsc, PhysRevD.101.063511, Liu_2020, KASE2020135400, PhysRevD.101.043531, sym12030481, Amendola_2020, Benisty:2020nql, Yang_2020, Barrow:2019jlm, Silva:2024ift, Shah:2022cji, deSouza:2024sfl, Halder:2024uao, Shah:2024rme, RS2025l}. Beyond these phenomenological and scalar–tensor theories, several recent studies have developed DM-DE interaction models based directly on particle physics and string-theoretic frameworks \cite{Khoury:2025txd,Bedroya:2025fwh}. These works provide concrete examples of theoretically motivated interactions that can naturally lead to evolving dark energy behavior, including trends broadly consistent with the departures from $\Lambda$CDM suggested by recent DESI measurements.

In this work, we focus on one of the most prominent alternatives to $\Lambda$CDM: the Chameleon dark energy model \cite{PhysRevLett.93.171104}. This model exhibits a particularly intriguing feature— that allows the dark energy equation of state parameter to drop below the phantom divide ($w < -1$) without violating the null energy condition or leading to catastrophic instabilities of ghost fields in the cosmic evolution \cite{Huey_2006,_tefan_i__2004,PhysRevD.73.083509}. The core dynamics of this model are governed by an interaction between dark matter and scalar-field dark energy, a coupling that emerges naturally from string-theory compactifications \cite{Damour_1994}. In this setup, energy flows from the dark energy field to the dark matter sector, modifying the expected dilution behavior of dark matter over cosmic time. For an observer unaware of such an interaction and interpreting the dark matter evolution within a $\Lambda$CDM-like framework, this energy exchange manifests as an apparent phantom behavior of the dark energy at earlier times.  
 
This work presents a detailed numerical investigation of the Chameleon dark energy and its implications for cosmology. In contrast to earlier work \cite{Boriero_2015}, where an approximate analytical solution of the scalar field was used, we solve the Klein-Gordon equation numerically by employing a shooting algorithm to accurately determine the initial condition of the scalar field. This more precise approach allows us to capture the field dynamics across all relevant redshifts, ensuring a consistent treatment throughout cosmic history. Previous works have also explored similar types of models and found a preference for non-zero coupling strength in the dark sector \cite{Gomez-Valent:2020mqn, Gomez-Valent:2022bku, Goh:2022gxo}. Interestingly, we observe that the combined datasets of Planck, DESI DR2 BAO, and Pantheon+ datasets prefer a higher value of the coupling strength compared to the results of previous analyses. Recently, sign-changing interacting dark energy models, where the flow of energy direction from dark matter to dark energy is reversed at a certain redshift, have also been able to constrain the interaction strength in the dark sector \cite{PhysRevD.111.043531}. However, our model, which naturally arises from string compactifications, constrains the coupling strength with a preference for higher non-zero coupling strength without any sign of change in dark sector coupling. This intriguing result opens the door to further explorations, in particular, the effect on non-linear scales, which alter the history of structure formation due to the violation of the equivalence principle. With the advent of upcoming cosmological surveys capable of probing cosmic structures with unprecedented precision, the observational imprint of this interaction could be identified with greater statistical significance. Moreover, a non-zero coupling is expected to generate a distinct signature in the differential infall of dark matter and baryons into virialized halos \cite{Kesden_2006,Keselman_2009,Secco_2018}. These effects will be systematically explored in our future work using dedicated N-body simulations, which will allow us to quantify the impact of the fifth force on structure formation in the non-linear regime.

The plan of the paper is as follows. In section \ref{sec:level2}, we briefly describe the chameleon model and its effect on cosmology. In section \ref{s3}, we elaborate on the numerical implementation of this model in the Boltzmann code \texttt{CLASS}\cite{lesgourgues2011cosmic, DiegoBlas_2011}. We present the method of our Markov Chain Monte Carlo (MCMC) analysis with different datasets in section \ref{s4} and discuss the result in section \ref{s5}. We conclude in section \ref{s6}.

\section{\label{sec:level2} The Chameleon Model}

This work presents a comprehensive investigation of a Yukawa-type interaction between dark matter and a quintessential scalar field functioning as dark energy. Such a light scalar degree of freedom can naturally emerge from string theory compactifications, enabling its coupling to the matter sector \cite{Damour_1994}. To remain consistent with the tight constraints from solar system gravity tests \cite{Will_2001,PhysRevLett.98.021101}, the Chameleon mechanism typically restricts this interaction to the dark matter component alone \cite{PhysRevLett.93.171104, PhysRevD.69.044026, PhysRevD.70.123518, BRAX2006441, COPELAND_2006, PhysRevLett.109.241301}. As a result, the scalar field dynamically acquires a mass that depends on the local dark matter density, giving rise to a screening effect that renders the field's influence negligible in high-density environments \cite{PhysRevD.73.083509,PhysRevLett.109.241301}. 

In this paper, we follow the model developed in \cite{PhysRevD.73.083509}, where the quintessential scalar field, $\phi$, couples to dark matter, $\psi$, through a Yukawa-type coupling of the form $f(\phi) \overline{\psi} \psi$. The dynamics of this field are governed by a modified Klein-Gordon equation, which includes an additional source term proportional to the dark matter energy density and the derivative of the coupling function.
\begin{equation}
    \phi^{\prime\prime} + 3H\phi^{\prime}= -V_{,\phi}(\phi) - \frac{\rho_{\rm dm}^{(0)}}{a^3} \  \frac{f_{,\phi}(\phi)}{f(\phi_0)}. 
    \label{eq_phi}
\end{equation}
Here $\phi^{\prime}$ denotes the derivative with respect to standard time. Consequently, the scalar field evolves under an effective potential that includes both the self-interaction potential and an interaction-induced term, which depends on the background dark matter density,
\begin{equation}
    V_{\text{eff}}(\phi) = V(\phi) + \frac{\rho_{\rm dm}^{(0)}}{a^3} \frac{f(\phi)}{f(\phi_0)},
    \label{eq_veff}
\end{equation}
Here, $\phi_0$ denotes the position of the scalar field in the present epoch. In this context, we adopt the following form of the coupling function, motivated by string theory compactifications \cite{Damour_1994},
\begin{equation}
    f(\phi) = \exp\left(\frac{\beta\phi}{\sqrt{8\pi }M_{\rm Pl}}\right),
    \label{eq_coupling}
\end{equation}

where $M_{\rm pl}$ is the reduced Planck mass. Due to the interaction with dark energy, the mass of dark matter becomes a function of the scalar field, causing its energy density to evolve as $\rho_{\rm dm}\sim f(\phi)/a^3$, which is slower than the standard $\Lambda$CDM evolution of $\sim a^{-3}$, since, similar to \cite{PhysRevD.73.083509}, we assume that $f(\phi)$ monotonically increases at low redshifts, when dark energy effects become prominent. The corresponding Friedmann equation becomes,

\begin{equation}
    3H^2 M_{\rm Pl}^2
    = \frac{\rho_{\rm dm}^0}{a^3}\frac{f(\phi)}{f(\phi_0)}
    + \rho_\phi + \rho_{\rm other}
\end{equation}

where $\rho_\phi= \phi^{\prime 2}/2 + V(\phi)$ and $\rho_{\rm other}= \rho_{\gamma}+ \rho_{\nu}+\rho_{\rm b}$. $\rho_{\rm other}$ represents the total energy density of photons, neutrinos, and baryons. These components redshift in the same way as in the standard $\Lambda$CDM model, since they are not affected by the coupling between dark matter and dark energy.

Importantly, an observer who is unaware of this interaction—and assumes that dark matter follows its standard evolution—would infer an effective non-interacting dark energy, fully described by an effective equation of state, $w_{\rm eff}$ and follows the following conservation equation,

\begin{equation}\label{eq:de_eff_cont}
    \frac{d\rho_{\rm de}^{\rm eff}}{dt}=-3H(1+w_{\rm eff})\rho_{\rm de}^{\rm eff}
\end{equation}

Correspondingly, the observer assumed the dark matter to be standard Cold Dark Matter (CDM), which results in the following Friedmann equation,

\begin{equation}\label{eq:eff_friedmann}
    3H^2 M_{\rm Pl}^2
    = \frac{\rho_{\rm dm}^{(0)}}{a^3}
    + \rho_{\rm de}^{\rm eff} + \rho_{\rm other}
\end{equation}

Due to the observer's assumption, this model would give rise to the effective dark energy density,

\begin{equation}\label{eq:rho_de_eff}
    \rho_{\rm de}^{\rm eff}=\frac{\rho_{\rm dm}^{(0)}}{a^3}\left[\frac{f(\phi)}{f(\phi_0)}-1\right] + \rho_\phi\
\end{equation}

Taking time derivative of the equation \ref{eq:rho_de_eff} and comparing it with equation \ref{eq:de_eff_cont} we get,

\begin{equation}\label{eq:weff_wphi}
    1+w_{\rm eff} = \frac{1}{\rho_{\rm de}^{\rm eff}}\left\{\left[\frac{f(\phi)}{f(\phi_0)}-1\right]\frac{\rho_{\rm dm}^{(0)}}{a^3} + (1+w_\phi)\rho_\phi\right\}
\end{equation}

where, the equation of state of the true scalar field dark energy,
\begin{equation}\label{eq:w_phi}
w_{\phi} = \left(\phi^{\prime 2}/2 - V(\phi)\right)/\left(\phi^{\prime 2}/2 + V(\phi)\right) ~.
\end{equation}

From equation \ref{eq:weff_wphi}, we find that $w_\phi$ and $w_{\rm eff}$ are connected through the following relation,

\begin{equation}\label{eq:w_eff}
  w_{\rm eff} = \frac{w_\phi}{1 - x} \quad \text{where,} \quad x = - \frac{\rho_{\rm dm}^0}{a^3 \rho_\phi} \left(\frac{f(\phi)}{f(\phi_0)} - 1\right),
\end{equation}

Although the quintessential scalar field itself does not exhibit any phantom behavior—remaining within the theoretical bounds of Null Energy Condition $-1 \leq w_{\phi} \leq -1/3$—an observer agnostic about the underlying dark sector interaction would still infer an effective equation of state that crosses into the phantom regime. This occurs because the parameter $x$ remains strictly positive for a monotonically increasing coupling function, thereby amplifying the deviation between $w_{\phi}$ and $w_{\rm eff}$. Such a mechanism could potentially account for the mild preference for phantom-like dark energy observed in several independent analyses \cite{PhysRevD.103.083533,D'Amico_2021,PhysRevD.99.123505}, and may serve as indirect evidence for an interaction within the dark sector.

We use the same runaway-type self-interaction potential given in \cite{PhysRevD.73.083509}, 
 \begin{equation}
    V(\phi) = M^4 \left( \frac{M_{Pl}}{\phi} \right)^{\alpha}.
    \label{eq_vphi}
\end{equation}
Here, $M$ represents the characteristic energy scale associated with the scalar field, while $\alpha$ denotes the steepness of the potential. The evolution of the quintessential scalar field follows a thawing scenario, wherein the field remains nearly frozen at its initial value at earlier epochs due to strong Hubble friction, and gradually begins to evolve as the Hubble parameter decreases. This evolution drives the field toward the dynamically evolving minimum of the effective potential. In addition to this, we impose an important condition that the scalar field must settle at the present-day minimum of the effective potential with vanishing kinetic energy. This ensures that the total contribution to the dark energy density today arises purely from the self-interaction potential. To implement this condition consistently, we normalize the self-interaction potential in the following way:

 \begin{equation} \label{eq:v_phi_norm}
     V(\phi)=3 H_0^2 M_{\rm pl}^2 \Omega_{\rm de}^0 \left(\frac{\phi_0}{\phi}\right)^\alpha
 \end{equation}

In our work, we use the analytical expression obtained in \cite{PhysRevD.73.083509} through minimizing the effective potential,

\begin{equation} \label{eq:phi0}
    \frac{\phi_0}{\sqrt{8\pi }M_{\rm Pl}}=\frac{\alpha}{\beta} \frac{\Omega_{\rm de}^0}{\Omega_{\rm dm}^0}
\end{equation}

Previous observations from Planck, BAO datasets (such as BOSS and eBOSS), and various supernova compilations have consistently pointed toward a present-day dark energy equation of state close to $w_{\rm de} = -1.0$, with strong statistical significance \cite{2020,2022MNRAS.512.5657S,Sevilla_Noarbe_2021}. As demonstrated in \cite{PhysRevD.73.083509}, the effective equation of state in the Chameleon framework naturally evolves toward $w_{\rm eff} \to -1$ as $z \to 0$, remaining compatible with these constraints. However, recent results from the DESI collaboration have provided strong evidence for evolving dark energy, thereby generating tension with the standard $\Lambda$CDM scenario \cite{2025desidr2}. It is important to note that this tension arises only when an evolving dark energy component is assumed in the fiducial cosmology used for the analysis. Moreover, DESI's analysis reveals a crossing of the phantom divide in the reconstructed dark energy equation of state around redshift $z \sim 0.5$ and $w_{\rm de} > -1.0$  today, as observed across several model-independent reconstruction techniques \cite{lodha2025}. While the statistical significance of this phantom crossing remains modest, a recent study has demonstrated an alternative realization of both of these behavior within the Chameleon framework \cite{chakraborty2025desi}. However, their approach differs substantially from the present work, particularly in the assumed scalar field mass and in allowing a non-zero kinetic energy for the field at the current epoch. Notably, in this work, the scalar field rolls down the effective potential after overcoming the Hubble friction and settles down eventually to the minima of the effective potential at the present epoch, leading to the dark energy equation of state, $w_{\rm de}$ close to $-1$. This eventually limits the model to the extent it can explain the DESI observation.

Due to the presence of an additional interaction in the dark sector, the perturbation equations will be modified accordingly. The equations take the following form in the synchronous gauge.

\begin{equation} 
\begin{aligned}
\dot{\delta}_{\rm dm}&=-\left( \frac{\theta_{\rm dm}}{a} + 
\frac{\dot{\tilde{h}}}{2} \right) + \frac{{\beta}}{\sqrt{8\pi}M_{\rm Pl}} \delta \dot{\phi} \ , \\
\dot{\theta}_{\rm dm} &= -H\theta_{\rm dm}+\frac{{\beta}}{\sqrt{8\pi}M_{\rm Pl}}\left( \frac{k^2}{a}\delta\phi -\dot{\phi}\theta_{\rm dm} \right) \ , \label{eq:coldvelocity}
\end{aligned}
\end{equation} 
Also, the perturbed Klein-Gordon equation becomes,
\begin{equation} 
\ddot{\delta \phi} + 3H\dot{\delta \phi}+\left( \frac{k^2}{a^2} +V_{,\phi\phi} \right)\delta\phi + \frac{1}{2}\dot{\tilde{h}}\dot{\phi}=-\frac{{\beta}}{\sqrt{8\pi}M_{\rm Pl}} \rho_{\rm dm}\delta_{\rm dm} \ ,  \label{eq:scalarperturbations}
\end{equation}  
where $\tilde{h} \equiv \tilde{h}^i_{\ i}$ is the trace of metric perturbation~$\tilde{h}_{ij}$ in synchronous gauge.

Similarly, the perturbed Einstein equation determining the evolution of metric perturbation gets modified too in the following manner,
\begin{widetext}
\begin{align}
k^2 \eta - \tfrac{1}{2} a^2 H \dot{\tilde h}
  &= -\frac{a^2}{2 M_{\rm Pl}^2}
     \left(\sum_{i=\gamma,\nu,{\rm b},{\rm dm}} \rho_i \delta_i + \delta\rho_\phi \right),
     \label{eq:metricperturb} \\
k^2 \dot{\eta}
  &= \frac{a}{2 M_{\rm Pl}^2}
     \left(\sum_{i=\gamma,\nu,{\rm b},{\rm dm}} (\rho_i+P_i)\,\theta_i
           + a k^2 \dot{\phi}\,\delta\phi \right), \\
\ddot{\tilde h} + 3 H \dot{\tilde h} - 2 \frac{k^2}{a^2} \eta
  &= -\frac{3}{M_{\rm Pl}^2}
     \left(\sum_{i=\gamma,\nu,{\rm b},{\rm dm}} \delta P_i
           + \dot{\phi}\,\delta\dot{\phi} - V_{,\phi}\,\delta\phi \right).
     \label{eq:metricperturb2}
\end{align}
\end{widetext}

where $\delta \rho_\phi = \dot{\phi}\delta\dot{\phi}+V_{,\phi}\delta\phi$. From Eq.~\ref{eq_veff}, it is evident that the dark matter density-dependent term dominates in the early universe due to the very small values of the scale factor. The self-interaction potential becomes comparable only once the scale factor is sufficiently large, which occurs around the epoch when dark energy starts to dominate. At this stage, the scalar field is gradually attracted toward the effective minimum, leading to a monotonic increase in its value. This late-time roll of the scalar field not only alters the background expansion history but also induces a time variation in the gravitational potentials.

An important observational imprint of this effect arises through the late-time Integrated Sachs--Wolfe (ISW) contribution to the CMB anisotropies. In standard cosmology, the gravitational potentials remain nearly constant during matter domination, and the ISW contribution is negligible. Once the expansion history deviates from pure matter domination, the potentials begin to evolve, leading to an ISW signal at large angular scales in the CMB. In chameleon models, the scalar--matter coupling modifies both the background acceleration and the growth of structure, thereby altering the time evolution of the potentials compared to $\Lambda$CDM. Although the ISW signal is limited by cosmic variance, cross-correlations with large-scale structure surveys provide a complementary probe of this distinctive feature of chameleon cosmology. The late-time ISW effect in the chameleon cosmology has been extensively discussed in previous works \cite{Boriero_2015}.

In the next section, we present the procedure of implementing the model numerically in terms of both modified background and perturbation dynamics, and we present the results in Section \ref{s5} by putting constraints on the model parameters from various observational datasets.

\section{Model Implementation: Shooting Algorithm} \label{s3}

In this study, we modify the publicly available Boltzmann code \texttt{CLASS} \cite{lesgourgues2011cosmic, DiegoBlas_2011} to incorporate the background and perturbation equations relevant to the Chameleon dark energy model. In particular, we numerically solve the Klein-Gordon equation in conjunction with the Friedmann equations and the perturbed Einstein equations, following the formalism outlined in \cite{PhysRevD.78.083538}. A critical aspect of this implementation involves the specification of the initial scalar field value, $\phi_i$. An arbitrary choice of $\phi_i$ does not ensure that the scalar field will evolve to the correct present-day minimum of the effective potential, $\phi_0$. As a result, this mismatch can lead to an inaccurate estimation of the present-day dark energy density, attributed to an incorrect treatment of the field’s background evolution. 

To ensure that the scalar field accurately reproduces the observed dark energy density today, our objective is to solve the Klein-Gordon equation such that the field evolves toward the minimum of the effective potential, $\phi_0$ given in Eq. \ref{eq:phi0}, with negligible kinetic energy at the present epoch. Achieving this behavior requires a careful choice of the initial field value and its velocity. To this end, we implement an automated shooting algorithm within \texttt{CLASS}, designed to iteratively adjust the initial field value to match the analytically expected $\phi_0$ at $z = 0$. Starting from a trial configuration of the initial scalar field and a fixed initial velocity set to zero, the code evolves the background cosmology and checks whether the final field value aligns with the target $\phi_0$. If a deviation is detected, the algorithm updates the initial field position and repeats the integration until the field reaches the desired final value within a numerical accuracy of at least $10^{-5}$.

This procedure is performed for each combination of the standard six $\Lambda$CDM cosmological parameters, along with input values of $\alpha$ and $\beta$. We have explicitly verified that the field velocity remains negligibly small at the present epoch, ensuring that its contribution to the total energy density is insignificant. Consequently, the present-day dark energy density arises almost entirely from the normalized self-interaction potential, as defined in Eq.~\ref{eq:v_phi_norm}. This method allows us to maintain high numerical precision in the evaluation of the energy density and enhances the robustness and accuracy of our results compared to previous studies that relied on approximate field dynamics.

Given the extremely light mass of the scalar field, its evolution is initially suppressed by Hubble friction, effectively freezing the field at its starting value in the early universe. The field begins to roll only at late times, and its dynamics are highly sensitive to the choice of $\alpha$ and $\beta$, which determine the slope and shape of the effective potential. At early epochs, the contribution from the dark matter coupling dominates the effective potential due to the small scale factor, while the self-interaction potential becomes relevant only at late times. As the field rolls down the effective potential, it can overshoot the minimum, leading to oscillatory behavior around $\phi_0$. However, since the crossing of the minimum occurs at late times, the field does not acquire sufficient velocity to significantly affect the present-day energy budget. The detailed dynamics of the scalar field evolution, including the minima crossing and oscillations, will be discussed further in Section \ref{s5}.

\begin{table*}[ht]
\caption{\label{tab:prior} Prior ranges considered for MCMC analysis for the cosmological parameters of Chameleon and $\Lambda$CDM model.
}
\begin{ruledtabular}
\begin{tabular}{lcc}
Parameters & Prior range  \\[1ex]
\hline\rule{0pt}{1.2\normalbaselineskip}

$100\omega_b$ & $\mathcal{U}[0.50,10.0]$ \\[1ex]
$\omega_{\rm dm}\footnote{for $\Lambda$CDM model, the 'dm' subscript means the CDM, and for the Chameleon model, it means the dark matter coupled with dark energy considered in this paper.}$ & $\mathcal{U}[0.0010,0.99]$  \\[1ex]
$100\times\theta_s$ & $\mathcal{U}[0.50,10.00]$  \\[1ex]
$\ln{10^{10}A_s}$ & $\mathcal{U}[1.60,3.90]$  \\[1ex]
$n_s$ & $\mathcal{U}[0.80,1.20]$  \\[1ex]
$\tau_{reio}$ & $\mathcal{U}[0.04,0.80]$  \\[1ex]
$\alpha$ & $\mathcal{U}[0.00,16.00]$  \\[1ex]
$\beta$ & $\mathcal{U}[0.00,1.00]$  \\[1ex]
\end{tabular}
\end{ruledtabular}
\end{table*}

\begin{table*}
\caption{\label{t_a1} The mean (bestfit) $\pm 1\sigma$ error for the cosmological parameters constrained with the datasets Planck and Planck + SDSS BAO for the $\Lambda$CDM model and Chameleon model are shown in this table. The $1\sigma$ error is obtained at $68\%$ confidence level. The corresponding $\chi_{min}^2$ and $\Delta \chi_{min}^2$ values are also shown. Upper bounds are obtained from the marginalized distribution such that 95\% of the samples lie below the quoted value, corresponding to the 95\% credible level.}
\begin{ruledtabular}
\begin{tabular}{ccccc}
Dataset & \multicolumn{2}{c}{Planck} & \multicolumn{2}{c}{Planck + SDSS BAO} \\[1ex] \hline\rule{0pt}{1.2\normalbaselineskip}
Model & $\Lambda$CDM & Chameleon & $\Lambda$CDM & Chameleon \\[1ex] \hline
\rule{0pt}{1.2\normalbaselineskip}
$100\omega_b$ & 
$2.24(2.25)\pm 0.0154$ &
$2.24(2.23)^{+0.0153}_{-0.0157}$ & 
$2.24(2.24)\pm 0.0143$ &
$2.24(2.24)\pm 0.0140$ \\[1ex]
$\omega_{\rm dm}\footnote{for $\Lambda$CDM model, the 'dm' subscript means the CDM, and for the Chameleon model, it means the dark matter coupled with dark energy considered in this paper.}$ & 
$0.120(0.119)\pm 0.00140$ &
$0.119(0.119)^{+0.00183}_{-0.00160}$ & 
$0.119(0.120)\pm 0.00102$ &
$0.119(0.118)^{+0.00112}_{-0.00107}$ \\[1ex]
$100\times\theta_s$ &
$1.04(1.04)\pm 0.000301$ &
$1.04(1.04)^{+0.000303}_{-0.000310}$ &
$1.042(1.042)\pm 0.000285$ &
$1.04(1.04)\pm 0.000293$ \\[1ex]
$\ln{10^{10}A_s}$ & 
$3.05(3.04)^{+0.0157}_{-0.0168}$ &
$3.04(3.04)\pm 0.0160$ & 
$3.05(3.04)^{+0.0156}_{-0.0173}$ &
$3.04(3.04)^{+0.0162}_{-0.0167}$ \\[1ex]
$n_s$ & 
$0.965(0.966)^{+0.00450}_{-0.00458}$ &
$0.966(0.966)^{+0.00462}_{-0.00465}$ & 
$0.967(0.964)\pm 0.00388$ &
$0.966(0.968)\pm 0.00391$ \\[1ex]
$\tau_{reio}$ & 
$0.0545(0.0521)^{+0.00763}_{-0.00820}$ &
$0.0538(0.0537)^{+0.00776}_{-0.00804}$ & 
$0.0557(0.0524)^{+0.00740}_{-0.00829}$ &
$0.0542(0.0535)^{+0.00771}_{-0.00817}$ \\[1ex]
$\alpha$ &
$-$ &
$<9.83$ &
$-$ &
$<9.55$ \\[1ex]
$\beta$ & 
$-$ &
$0.222(0.159)^{+0.0975}_{-0.105}$ &
$-$ &
$0.219(0.210)^{+0.0949}_{-0.0809}$ \\[1ex] \hline\rule{0pt}{1.2\normalbaselineskip}
$H_0$ &
$67.3(67.7)^{+0.625}_{-0.635}$ &
$68.2(67.6)^{+0.776}_{-1.03}$ &
$67.7(67.7)^{+0.446}_{-0.470}$ &
$68.2(68.4)^{+0.538}_{-0.606}$ \\[1ex]
$\sigma_8$ &
$0.811(0.806)^{+0.00771}_{-0.00763}$ &
$0.824(0.813)^{+0.0104}_{-0.0143}$ &
$0.809(0.807)^{+0.00700}_{-0.00764}$ &
$0.823(0.832)^{+0.0103}_{-0.0137}$ \\[1ex] \hline
 \rule{0pt}{1.2\normalbaselineskip}
$\chi^2_{\rm min}$ &
$2772.09$ &
$2771.78$ &
$2777.08$ &
$2776.52$ \\[1ex] \hline
 \rule{0pt}{1.2\normalbaselineskip}
$\Delta \chi^2_{\rm min}$ &
$0$ &
$-0.31$ &
$0$ &
$-0.56$ \\
\end{tabular}
\end{ruledtabular}
\end{table*}

\begin{table*}
\caption{\label{t_a2} The mean (bestfit) $\pm 1\sigma$ error for the cosmological parameters constrained with the datasets Planck + SDSS BAO + Pantheon Plus and Planck + SDSS BAO + Pantheon Plus + SH$_0$ES for the $\Lambda$CDM model and Chameleon model are shown in this table. The $1\sigma$ error is obtained at $68\%$ confidence level. The corresponding $\chi_{min}^2$ and $\Delta \chi_{min}^2$ values are also shown. Upper bounds are obtained from the marginalized distribution such that 95\% of the samples lie below the quoted value, corresponding to the 95\% credible level.}
\begin{ruledtabular}
\begin{tabular}{ccccc}
Dataset & \multicolumn{2}{c}{Planck + SDSS BAO + Pantheon+} & \multicolumn{2}{c}{Planck + SDSS BAO + Pantheon+ + SH$_0$ES} \\[1ex] \hline\rule{0pt}{1.2\normalbaselineskip}
Model & $\Lambda$CDM & Chameleon & $\Lambda$CDM & Chameleon \\[1ex] \hline
\rule{0pt}{1.2\normalbaselineskip}
$100\omega_b$ &
$2.24(2.24)^{+0.0138}_{-0.0134}$ &
$2.24(2.24)^{+0.0137}_{-0.0140}$ & 
$2.26(2.25)^{+0.0135}_{-0.0132}$ &
$2.25(2.26)\pm 0.0141$ \\[1ex]
$\omega_{\rm dm}\footnote{for $\Lambda$CDM model, the 'dm' subscript means the CDM, and for the Chameleon model, it means the dark matter coupled with dark energy considered in this paper.}$ & 
$0.120(0.119)^{+0.000950}_{-0.000970}$ &
$0.119(0.120)^{+0.00103}_{-0.00101}$ &
$0.117(0.117)^{+0.000880}_{-0.000860}$ &
$0.117(0.117)\pm 0.000937$ \\[1ex]
$100\times\theta_s$ &
$1.04(1.04)^{+0.000300}_{-0.000260}$ &
$1.04(1.04)\pm 0.000291$ &
$1.04(1.04)^{+0.000280}_{-0.000270}$ &
$1.04(1.04)^{+0.000295}_{-0.000292}$ \\[1ex]
$\ln{10^{10}A_s}$ & 
$3.04(3.04)^{+0.0155}_{-0.0161}$ &
$3.04(3.04)^{+0.0158}_{-0.0163}$ &
$3.05(3.04)^{+0.0164}_{-0.0180}$ &
$3.04(3.06)^{+0.0158}_{-0.0169}$ \\[1ex]
$n_s$ & 
$0.966(0.965)^{+0.00363}_{-0.00379}$ &
$0.965(0.967)^{+0.00380}_{-0.00384}$ &
$0.972(0.973)^{+0.00374}_{-0.00364}$ &
$0.969(0.97)^{+0.00395}_{-0.00385}$ \\[1ex]
$\tau_{reio}$ & 
$0.0552(0.0513)^{+0.00750}_{-0.00776}$ &
$0.0536(0.0508)^{+0.00750}_{-0.00785}$ &
$0.0589(0.0574)^{+0.00752}_{-0.00903}$ &
$0.0550(0.0580)^{+0.00752}_{-0.00825}$ \\[1ex]
$\alpha$ &
$-$ &
$<8.89$ &
$-$ &
$<11.9$ \\[1ex]
$\beta$ & 
$-$ &
$0.197(0.154)^{+0.0859}_{-0.0815}$ &
$-$ &
$0.335(0.290)^{+0.0815}_{-0.0541}$ \\[1ex] \hline
\rule{0pt}{1.2\normalbaselineskip}
$H_0$ &
$67.5(67.8)^{+0.425}_{-0.430}$ &
$67.9(67.6)^{+0.484}_{-0.541}$ &
$68.6(68.6)^{+0.386}_{-0.398}$ &
$69.4(69.3)\pm 0.522$ \\[1ex]
$\sigma_8$ &
$0.810(0.803)^{+0.00726}_{-0.00689}$ &
$0.821(0.820)^{+0.00952}_{-0.0125}$ &
$0.804(0.803)^{+0.00727}_{-0.00778}$ &
$0.835(0.832)^{+0.0143}_{-0.0147}$ \\[1ex] \hline
\rule{0pt}{1.2\normalbaselineskip}
$\chi^2_{\rm min}$ &
$4189.92$ &
$4187.96$ &
$4104.19$ &
$4097.78$ \\[1ex] \hline
\rule{0pt}{1.2\normalbaselineskip}
$\Delta \chi^2_{\rm min}$ &
$0$ &
$-1.96$ &
$0$ &
$-6.41$ \\
\end{tabular}
\end{ruledtabular}
\end{table*}

\begin{table*}
\caption{\label{ta3} The mean (bestfit) $\pm 1\sigma$ error for the cosmological parameters constrained with the datasets Planck + DESI DR2 BAO + Pantheon Plus for the $\Lambda$CDM model and Chameleon model are shown in this table. The $1\sigma$ error is obtained at $68\%$ confidence level. The corresponding $\chi_{min}^2$ and $\Delta \chi_{min}^2$ values are also shown. Upper bounds are obtained from the marginalized distribution such that 95\% of the samples lie below the quoted value, corresponding to the 95\% credible level.}
\begin{ruledtabular}
\begin{tabular}{ccc}
 Dataset&\multicolumn{2}{c}{Planck + DESI DR2 BAO + Pantheon+}\\[1ex] \hline\rule{0pt}{1.2\normalbaselineskip}
 Parameter&$\Lambda$CDM&Chameleon \\[1ex] \hline
\rule{0pt}{1.2\normalbaselineskip}
$100\omega_b$&
$2.25(2.26)\pm 0.0131$&
$2.24(2.26)^{+0.0136}_{-0.0138}$ \\[1ex]
 $\omega_{\rm dm}\footnote{for $\Lambda$CDM model, the 'dm' subscript means the CDM, and for the Chameleon model, it means the dark matter coupled with dark energy considered in this paper.}$&
 $0.118(0.118)^{+0.000665}_{-0.000688}$&
 $0.118(0.117)^{+0.000677}_{-0.000673}$\\[1ex]
 $100\times\theta_s$&
 $1.04(1.04)\pm 0.000278$&
 $1.04(1.04)^{+0.000285}_{-0.000296}$ \\[1ex]
 $\ln{10^{10}A_s}$&
 $3.05(3.05)^{+0.0160}_{-0.0177}$&
 $3.04(3.04)^{+0.0161}_{-0.0166}$\\[1ex]
 $n_s$&
 $0.971(0.972)\pm 0.00342$&
 $0.968(0.971)\pm 0.00371$\\[1ex]
 $\tau_{reio}$&
 $0.0575(0.0580)^{+0.00764}_{-0.00845}$&
 $0.0550(0.0538)^{+0.00760}_{-0.00815}$ \\[1ex]
 $\alpha$&
 $-$&
 $<10.1$ \\[1ex]
 $\beta$&
 $-$&
 $0.256(0.23)^{+0.0924}_{-0.0629}$ \\[1ex] \hline\rule{0pt}{1.2\normalbaselineskip}
 $H_0$&
 $68.4(67.4)\pm 0.302$&
 $68.7(69.0)^{+0.348}_{-0.366}$ \\[1ex]
 $\sigma_8$&
 $0.805(0.832)^{+0.00687}_{-0.00751}$&
 $0.825(0.823)^{+0.0121}_{-0.0143}$ \\[1ex]
 \hline
 \rule{0pt}{1.2\normalbaselineskip}
 $\chi^2_{\rm min}$ &
 $4200.94$&
 $4196.19$ \\[1ex] \hline
\rule{0pt}{1.2\normalbaselineskip}
$\Delta \chi^2_{\rm min}$& 
$0$& 
$-4.75$ \\
\end{tabular}
\end{ruledtabular}
\end{table*}

\begin{table*}[ht]
\caption{\label{tab:t_aic} Comparison of $\Delta \chi^2_{\rm min}$ and $\Delta$AIC per experiment for Chameleon and $\Lambda$CDM models.
}
\begin{ruledtabular}
\begin{tabular}{lccc}
\textbf{Dataset} & \multicolumn{2}{c}{Chameleon}  \\[1ex]
\hline\rule{0pt}{1.2\normalbaselineskip}
 & $\Delta \chi^2_{\rm min}$ & $\Delta$AIC  \\[1ex]
\hline\rule{0pt}{1.2\normalbaselineskip}

Planck & -0.31 & +3.69 \\[1ex]
Planck+ SDSS BAO & -0.56 & +3.44  \\[1ex]
Planck + SDSS BAO + Pantheon+ & -1.96 & +2.04  \\[1ex]
Planck + DESI DR2 BAO + Pantheon+ & -4.75 & -0.75  \\[1ex]
Planck + SDSS BAO + Pantheon+ + SH$_0$ES & -6.41 & -2.41  \\[1ex]

\end{tabular}
\end{ruledtabular}
\end{table*}

\section{Method} \label{s4}

We modify the Boltzmann code \texttt{CLASS} to incorporate the background and perturbed Einstein equations specific to the Chameleon dark energy model, following the formulation presented in \cite{PhysRevD.78.083538}. This allows us to compute both the cosmic microwave background (CMB) temperature and polarization anisotropy power spectra, as well as the linear matter power spectrum within the Chameleon framework.

To constrain the model parameters, we perform a comprehensive MCMC analysis using the following cosmological datasets:

\begin{itemize}
\item[$\blacktriangleright$] \textbf{Planck \hspace{0.5mm}2018:}\hspace{2mm}High-$\ell$ temperature (TT), temperature-polarization (TE), and polarization (EE) power spectra, along with low-$\ell$ TT and EE measurements \cite{2020}.
\vspace{2mm}
\item[$\blacktriangleright$]  \textbf{SDSS BAO:} Baryon Acoustic Oscillation measurements from SDSS-III BOSS DR12 galaxy sample at redshifts $z = 0.38,\ 0.51,\ 0.61$ \cite{Alam_2017}, 6dF Galaxy Survey (6dFGS) at $z = 0.106$ \cite{Beutler_2011}, and SDSS DR7 Main Galaxy Sample at $z = 0.15$ \cite{Ross_2015}. 
\vspace{2mm}
\item[$\blacktriangleright$]\textbf{DESI DR2 BAO:} Baryon Acoustic Oscillation measurements from DESI’s second data release \cite{2025desidr2}, including tracers from galaxies, quasars, and the Lyman-$\alpha$ forest. The data provide both isotropic and anisotropic constraints across the redshift range $0.295 \leq z \leq 2.330$, divided into nine redshift bins as detailed in Table IV of \cite{2025desidr2}.
\vspace{2mm}
\item[$\blacktriangleright$] \textbf{Pantheon+ :} Type Ia supernovae measurements covering a redshift range of $z = 0.001$ to $2.26$ \cite{Brout_2022}. 
\vspace{2mm}
\item[$\blacktriangleright$]  \textbf{SH$_0$ES Cepheid sample:} Local distance ladder measurements using Cepheid variables in host galaxies of nearby Type Ia supernovae with $z \leq 0.01$ \cite{Riess_2022}.
\end{itemize}

Our baseline cosmological model includes the six standard $\Lambda$CDM parameters: [$\omega_b$, $\omega_{dm}$, $\theta_s$, $\ln(10^{10}A_s)$, $n_s$, and $\tau_{reio}$], in addition to the two Chameleon model parameters, $\alpha$ and $\beta$. Throughout our MCMC analysis, we consider uniform priors for all the cosmological parameters, and the corresponding prior range for the model parameters is presented in Table \ref{tab:prior}.

The MCMC analysis is performed using the publicly available package \texttt{MontePython-v3} \cite{brinckmann2018montepython}, which employs the Metropolis-Hastings algorithm and interfaces directly with the modified version of \texttt{CLASS}. Parameter sampling efficiency is improved using Cholesky decomposition, which separates fast and slow directions in the parameter space \cite{Lewis_2000}. The convergence of the MCMC chains is monitored using the Gelman-Rubin criterion, requiring $R-1 < 0.01$ for all parameters \cite{gelman1992inference}. Additionally, we obtain the minimum $\chi^2$ values using the Sequential Least Squares Programming (SLSQP) optimizer implemented within \texttt{MontePython}.

To further quantify the model preference, we calculate the AIC, which penalizes models with additional free parameters. The AIC difference between the Chameleon model and $\Lambda$CDM is given by \cite{1100705}: 
\begin{equation}\label{eq:AIC}
    \Delta \text{AIC}=  \chi^2_{\rm min, Model} -  \chi^2_{\rm {min},\Lambda \rm CDM} + 2(N_{\rm Model}-N_{\Lambda \rm CDM})\,.
\end{equation} 
Here, the $\Lambda$CDM model is adopted as the reference. A negative value of $\Delta \text{AIC}$ indicates that the Chameleon model provides a superior fit to the observational data relative to $\Lambda$CDM for a given combination of datasets; in contrast, positive values suggest an inferior fit. For this work we adopt the following criterion \cite{mcgb-ntwr}: In favor of the model, $\Delta\mathrm{AIC}\in[-2,0)$ indicates weak evidence; $\Delta\mathrm{AIC}\in[-6,-2)$ indicates positive evidence; $\Delta\mathrm{AIC}\in[-10,-6)$ indicates strong evidence; and $\Delta\mathrm{AIC}< -10$ very strong evidence. Conversely, against the model $\Delta\mathrm{AIC}\in(0,2]$ indicates weak evidence; $\Delta\mathrm{AIC}\in(2,6]$ indicates positive evidence; $\Delta\mathrm{AIC}\in(6,10]$ indicates strong evidence; and $\Delta\mathrm{AIC}>10$ very strong evidence.

\begin{figure*}[t]
\includegraphics[width=1.0\textwidth]{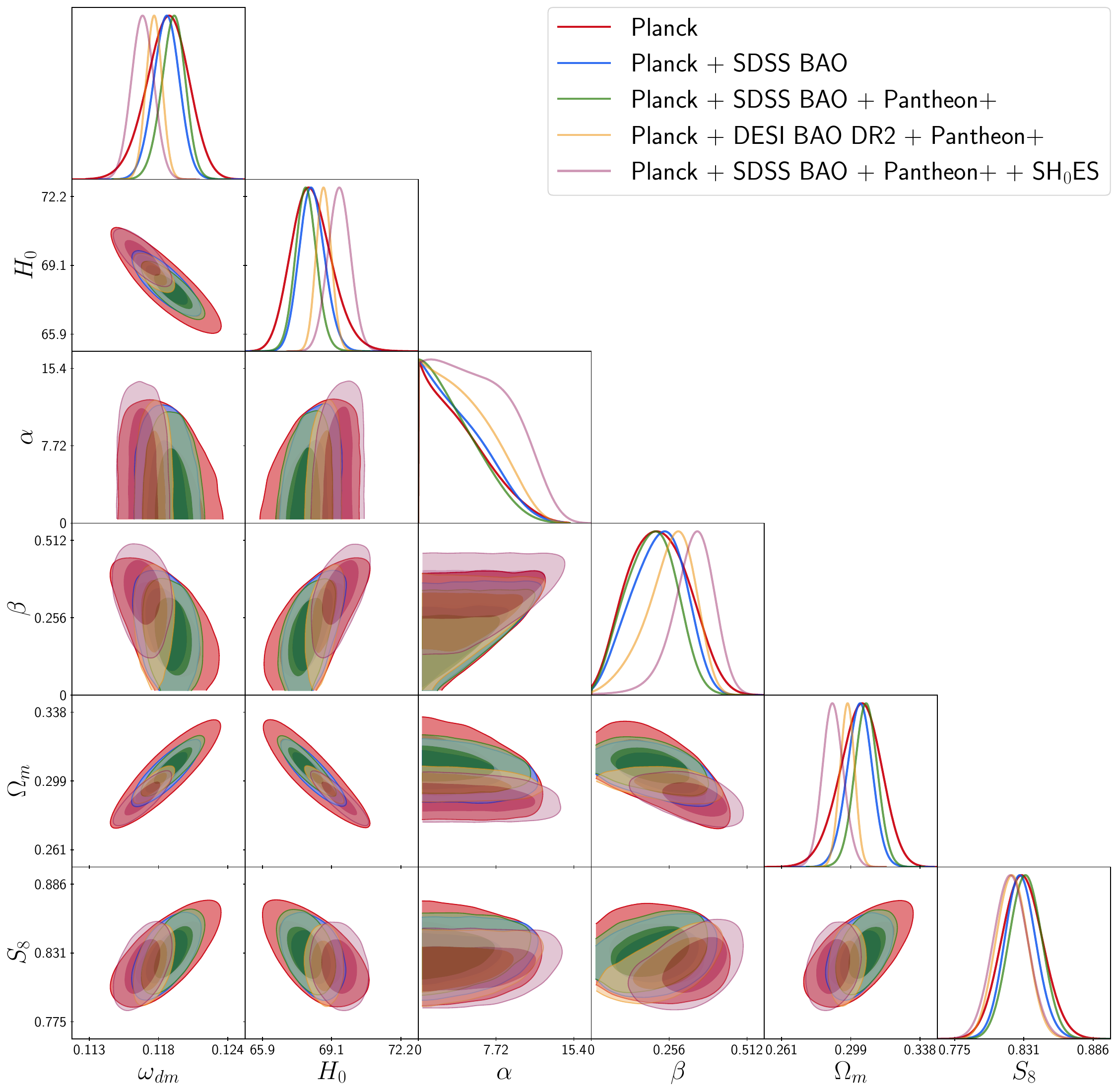}
\caption{Two-dimensional marginalized posterior distributions of the model 
parameters ($\omega_{\rm dm}$, $H_0$, $\alpha$, $\beta$, $\Omega_m$, and $S_8$) 
obtained from our MCMC analysis. The contours correspond to the 68\% and 95\% 
confidence regions. Different colors represent different dataset combinations: 
Planck-only, Planck+SDSS BAO, Planck+SDSS BAO+Pantheon+, 
Planck+DESI DR2 BAO+Pantheon+, and 
Planck+SDSS BAO+Pantheon+ + S$H_0$ES.}

\label{f_a3}
\end{figure*}

\begin{figure*}[ht]
\includegraphics[scale=0.38]{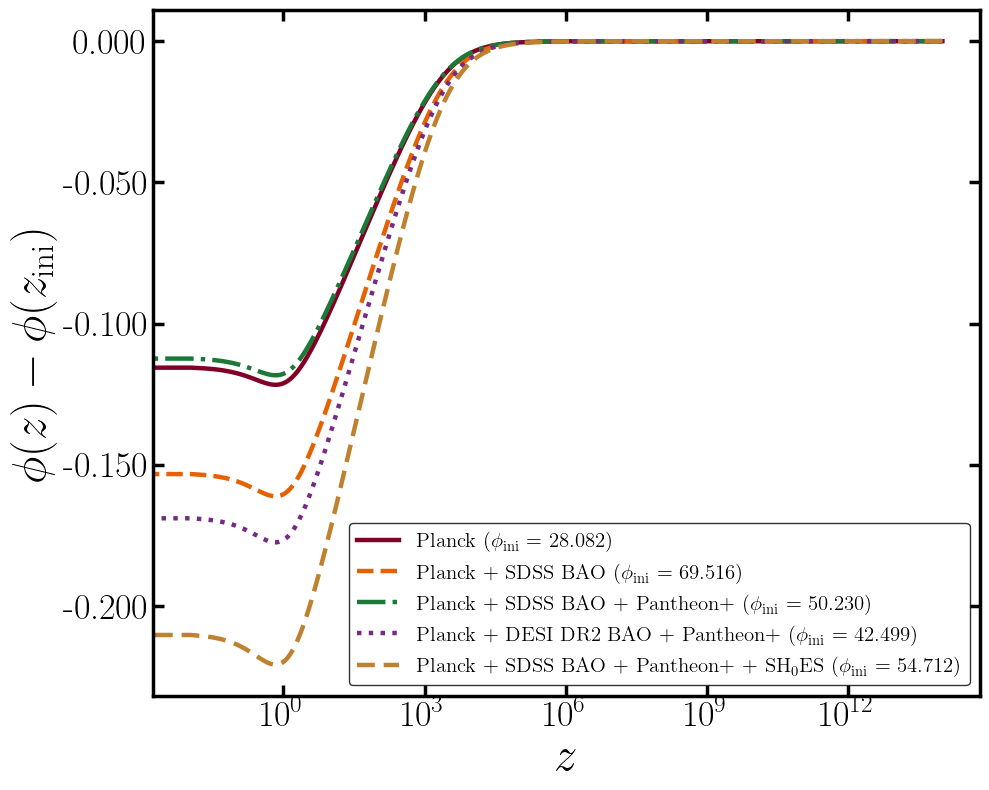}~
\includegraphics[scale=0.38]{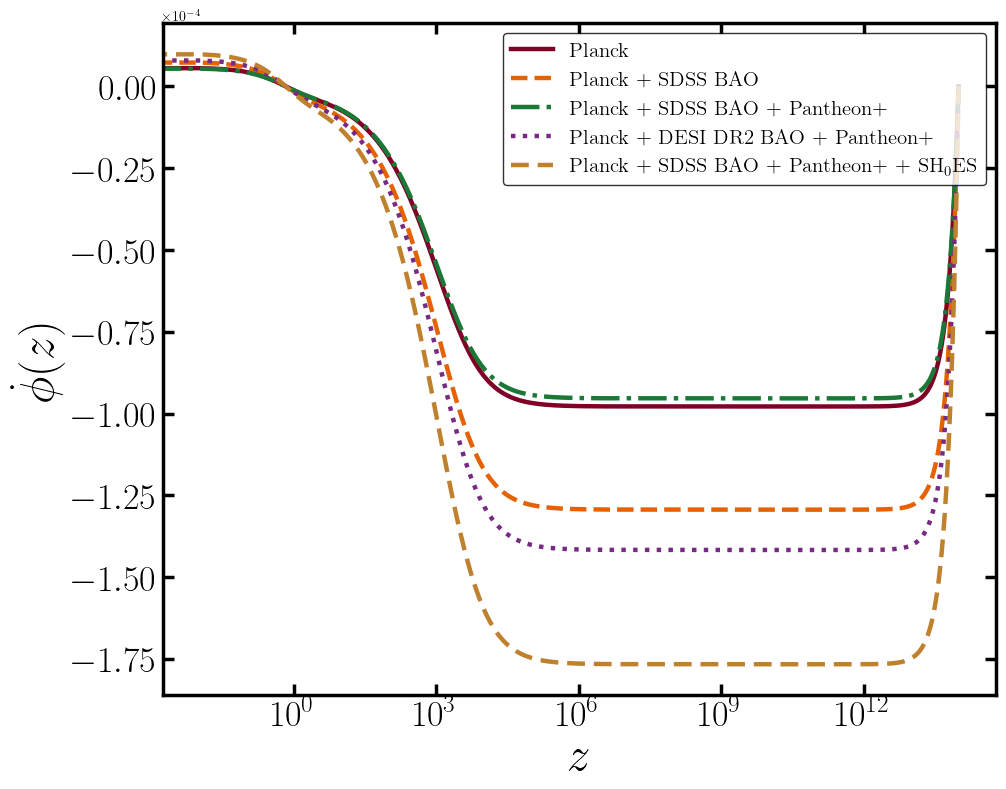}
\caption{\label{fig:scalar_dynamics}The scalar field dynamics with redshift is plotted here for best-fit values obtained from MCMC analysis for different combinations of datasets. The left panel shows the field value normalized to its initial value at the initial redshift, and the corresponding initial value is provided in the legend box. The right panel shows the corresponding dynamics of field velocity for the same set of best-fit values. Here $\dot{\phi}$ means derivative with respect to conformal time.}
\end{figure*}

\begin{figure*}
\centering
\includegraphics[scale=0.38]{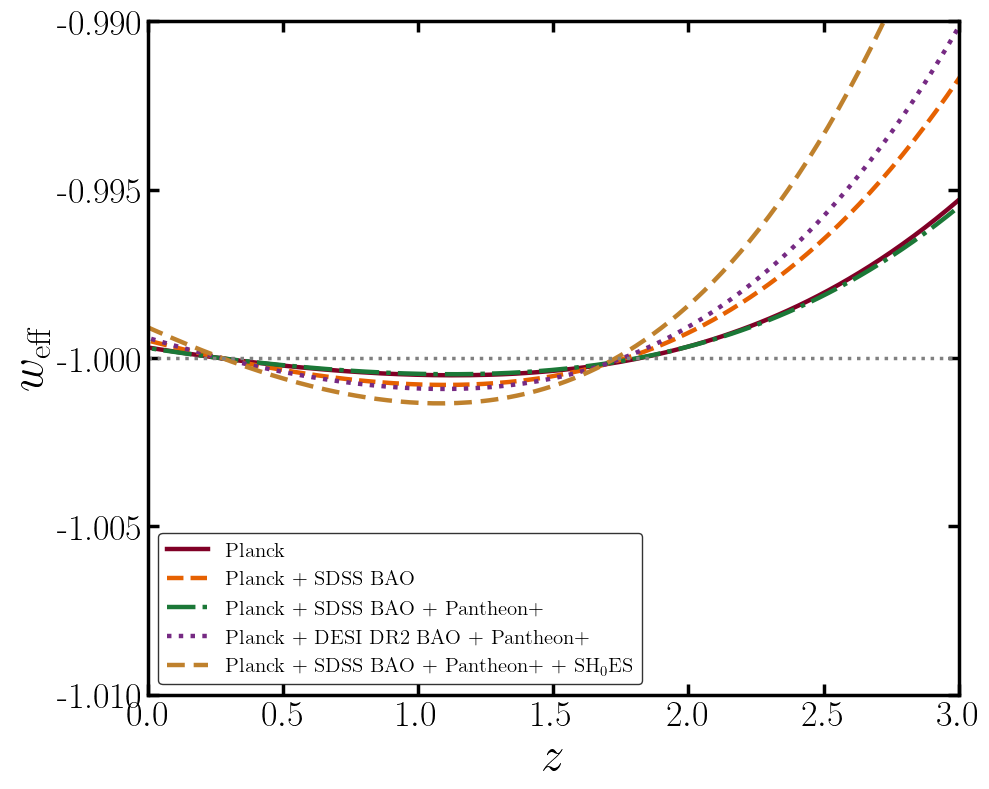}
\caption{\label{fig:w_eff_plot}The effective dark energy equation of state for the chameleon model is plotted with redshift for best-fit values obtained from MCMC analysis for different combinations of datasets. It highlights a phantom crossing at low redshift.}
\end{figure*}

\section{Results} \label{s5}

In this section, we present the results of our MCMC analysis of the model. The results are given in Tables~\ref{t_a1}, \ref{t_a2}, and \ref{ta3}, and illustrated in Figure~\ref{f_a3}, while the scalar field evolution and effective dark energy equation of state from the best-fit parameters are shown in Figures~\ref{fig:scalar_dynamics} and \ref{fig:w_eff_plot}. We also compare our model relative to $\Lambda$CDM statistically using the Akaike Information Criterion (AIC) in Table~\ref{tab:t_aic} for each dataset combination used for the analysis.

Figure \ref{f_a3} displays the one-dimensional marginalized posteriors and two-dimensional confidence contours for the model parameters. Notably, we find that while Planck data alone is capable of constraining $\beta$, our model does not show higher statistical significance compared to $\Lambda$CDM when using Planck-only data, as evident from such a low $\Delta \chi^2_{\rm min}$, presented in Table \ref{tab:t_aic}. We observe that the mean value of $\beta$ remains close to $\sim 0.200$ with the inclusion of additional datasets such as SDSS BAO and Pantheon+, while the constraints tighten progressively as more data are incorporated. Remarkably, when substituting the BAO measurements with those from DESI DR2, it yields a higher value of interaction strength, $\beta \sim 0.256$, and significantly tighter bounds relative to earlier results due to the higher sensitivity and precision in measurement of the BAO scalae by the DESI collaboration. Also, we will later show that our model also shows features of an evolving dark energy with a phantom crossing like DESI, as shown in Figure \ref{fig:w_eff_plot}. However, as mentioned in Section \ref{sec:level2}, departure of the effective dark energy equation of state from a cosmological constant remains tiny.

Furthermore, the inclusion of the DESI DR2 dataset leads to a modest upward shift in the Hubble constant, with $H_0 \sim 68.7$, compared to $H_0 \sim 68.0$ obtained from other datasets excluding DESI and closely aligned with the Planck baseline. When the full combination of Planck, SDSS BAO, Pantheon+, and SH$_0$ES data is considered, the model exhibits a preference for even higher dark sector interaction, with $\beta \sim 0.335$, along with the most stringent constraint on this parameter, as a result of the incredible precision of late-time dynamics from measurements of SH$_0$ES collaboration. The inclusion of SH$_0$ES also results in a further increase in the inferred Hubble constant, a trend that is clearly reflected in our analysis.   

Figure \ref{f_a3} also provides an upper bound on the parameter $\alpha$, which characterizes the slope of the scalar field's self-interaction potential. For the theoretical consistency of the model, $\alpha$ must remain strictly positive. As a result, the MCMC analysis can only yield an upper bound on this parameter, rather than tightly constraining it. A very small value of $\alpha$ corresponds to an extremely flat potential, implying the absence of a well-defined minimum near the present epoch. In such cases, the scalar field continues to evolve slowly with a decreasing kinetic energy, never settling into a stable configuration. On the other hand, a large value of $\alpha$ results in a steeper potential, causing the field to become increasingly confined and suppressing its evolution at late times. Therefore, an upper bound on $\alpha$ is an expected outcome, which is indeed observed in our analysis using Planck, SDSS BAO, and Pantheon+ datasets. 

Interestingly, the inclusion of the DESI DR2 BAO dataset slightly relaxes the bound on $\alpha$. This trend becomes more pronounced when the SH$_0$ES dataset is included, which further weakens the upper constraint on $\alpha$. It is also noteworthy that both DESI DR2 and SH$_0$ES datasets tend to favor slightly lower values of $\omega_{\rm dm}$, compared to the combinations without them. This can be attributed to the modified evolution of the dark matter energy density in the Chameleon framework, given by $\rho_{\rm dm} = \frac{\rho_{\rm dm}^{(0)}}{a^3} \  \frac{f_{,\phi}(\phi)}{f(\phi_0)}$, where the exponential coupling $f(\phi)$ depends on the interaction strength $\beta$ as defined in Eq. \ref{eq_coupling}. Since these datasets prefer higher values of both $\beta$ and $H_0$, the resulting evolution of $\rho_{\rm dm}$ leads to a lower preferred value of $\omega_{\rm dm}$. Consequently, the total matter density $\Omega_m$ is slightly reduced, resulting in a mild downward shift in the preferred value of $S_8$ relative to other dataset combinations. A more detailed investigation into the growth of cosmic structures—and the corresponding inference of the matter fluctuation amplitude $\sigma_8$—requires a dedicated analysis involving small-scale non-linear dynamics, which are best captured through N-body simulations. We leave this aspect of the analysis to future work.  

We also investigate the evolution of the scalar field and its dynamics by plotting the field value, its velocity, and the effective dark energy equation of state for the MCMC best-fit values obtained from various dataset combinations. These results are presented in Figures~\ref{fig:scalar_dynamics} and \ref{fig:w_eff_plot}. The field evolution is computed by numerically solving the Klein-Gordon equation \ref{eq_phi} using the shooting algorithm described in Section \ref{s3} to determine the appropriate initial field value that ensures consistency with present-day boundary conditions. 

The left panel of Figure~\ref{fig:scalar_dynamics} shows the evolution of the scalar field, normalized by subtracting its initial value at the initial epoch, while the right panel displays the corresponding evolution of the field velocity. As the plots indicate, the scalar field remains nearly frozen at its initial value during the early universe due to strong Hubble friction, with its velocity remaining close to zero. As the Hubble parameter gradually decreases over time, the field begins to roll.

During the early evolution, the scalar field experiences an effective potential dominated by the dark matter coupling term, which is enhanced by the small scale factor. This causes the field to roll down an exponentially shaped potential, leading to a monotonic decrease in its value. At later times, as the contribution from the self-interaction potential becomes comparable to the coupling term, a dynamically evolving minimum forms in the effective potential. The field accelerates toward this minimum, overshoots it due to its acquired velocity, and then begins to decelerate. After reaching a turning point—where its velocity momentarily vanishes—it reverses direction and begins to roll back down the potential.

This process leads to a damped oscillatory motion around the evolving minimum of the effective potential. The field’s velocity and position stabilize as it settles into the minimum with a negligible residual velocity, consistent with the boundary condition imposed at the present epoch. This behavior encapsulates the characteristic dynamics of the Chameleon mechanism and its late-time convergence to a potential-dominated dark energy state.

Using the scalar field dynamics obtained by numerically solving the Klein-Gordon equation, we compute the evolution of the effective dark energy equation of state, $w_{\rm eff}$, as defined in Eqs. \ref{eq:w_eff} \& \ref{eq:w_phi}. The results are presented in Figure \ref{fig:w_eff_plot}, which provides the evolution focused on the low-redshift regime, precisely where the Chameleon dark energy dynamics become observationally relevant.

In the context of recent evidence of dynamical dark energy from DESI observations, we analyze our Chameleon dark energy model using the DESI DR2 BAO dataset and find a preference for a non-zero dark sector coupling, with the interaction strength constrained to $\beta \sim 0.26$. Correspondingly, our model exhibits a crossing of the phantom divide in the effective dark energy equation of state around redshift $z \sim 0.5$, as illustrated in Figure~\ref{fig:w_eff_plot}. However, the evolution of $w_{\rm eff}$ remains very close cosmological constant at all relevant redshifts as a result of the assumptions of our model, mentioned in Section \ref{sec:level2}, which leads to $w_{\rm eff}$ close to $-1$ at present epoch. This conflicts with DESI observed preference of dark energy equation of state being higher than $-1$ today and explains the reduction of $\Delta \chi^2_{\rm min}$, as shown in Table \ref{ta3} is not as significant as obtained by DESI collaboration for $w_0 w_a$CDM model.

Although the statistical significance of the phantom crossing in the current DESI dataset remains moderate, the consistent emergence of this feature across independent analyses and within our framework suggests it may be more than a statistical fluctuation. Future DESI data—with improved redshift resolution and increased statistical power—will be essential in further probing the nature of dark energy. These forthcoming observations will allow for more stringent tests of interacting dark sector models like ours, potentially confirming the presence of long-range interactions or ruling out this class of models entirely.

To assess the statistical significance of the Chameleon model relative to the standard $\Lambda$CDM scenario, we compute the difference in the minimum chi-squared value, $\Delta \chi^2_{\rm min}$, for each dataset combination used in our MCMC analysis. The results are summarized in Table~\ref{tab:t_aic}. Across all combinations, we observe a reduction in $\chi^2_{\rm min}$ when the Chameleon model is used and when only Planck, SDSS BAO, and Pantheon+ datasets are considered; the amount of reduction remains consistent with other interacting dark energy models\cite{Giare:2024smz}. The most notable improvement occurs when the SH$0$ES dataset is included, yielding a maximum reduction of $\Delta \chi^2{\rm min} = -6.41$. A similarly significant reduction is observed when DESI DR2 BAO is used in place of SDSS BAO, with $\Delta \chi^2_{\rm min} = -4.75$, indicating a possible mild statistical preference for interaction in the dark sector in both cases of the combined datasets.

To statistically quantify the model preference, we calculate the $\Delta$AIC using equation \ref{eq:AIC} by comparing our model with $\Lambda$CDM for each dataset combination used in our analysis. The results are presented in Table \ref{tab:t_aic}, and we base our conclusion on the adopted criterion mentioned in section \ref{s4}. As shown in Table~\ref{tab:t_aic}, we get weak or positive evidence in terms of favoring the chameleon model relative to $\Lambda$CDM when either the DESI DR2 BAO or SH$_0$ES dataset is included, yielding $\Delta$AIC values of $-0.75$ and $-2.41$, respectively. These results provide statistically meaningful (weak/positive) preference for interaction in the dark sector, with the combination of Planck, SDSS BAO, Pantheon+, and SH$_0$ES offering the positive evidence. This dataset combination is particularly sensitive to late-time cosmological dynamics, making it well-suited to probe deviations from the $\Lambda$CDM baseline. Furthermore, replacing SDSS BAO with DESI DR2 BAO also leads to weak evidence of dark sector interaction, suggesting that DESI data may offer a mild indirect indication of such interactions in the dark sector.

\section{Conclusion} \label{s6}

In this work, we investigate the possibility of an interaction within the dark sector using current cosmological datasets. Specifically, we consider the Chameleon dark energy model, where a quintessential scalar field is coupled to dark matter via a Yukawa-type interaction motivated by string compactification. In the absence of knowledge about this coupling, an observer would consider an incorrect evolution of the dark energy density, resulting in an observation of an effective dark energy that exhibits phantom-like behavior. The scalar field evolves under an effective potential comprising two contributions: the self-interaction potential and the interaction term sourced by the dark matter density, together forming a dynamically evolving minimum. We further impose the condition that the scalar field must settle at this minimum with vanishing velocity at the present epoch. To implement this scenario, we numerically solve the Klein-Gordon equation using a shooting algorithm, ensuring consistency with our boundary conditions. Based on the resulting scalar field dynamics, we compute the effective dark energy equation of state, which exhibits phantom behavior at earlier epochs around $z\sim 2.0$, below which marks the onset of dark energy domination (around $z\sim 1.0$). 

We perform an MCMC analysis of the Chameleon model using various combinations of datasets, including Planck, SDSS BAO, DESI DR2 BAO, Pantheon+, and SH$0$ES. We find that all dataset combinations are able to fully constrain the interaction strength $\beta$ without requiring any sign-switching mechanism or additional exotic assumptions. We find higher non-zero interaction strength when incorporating the DESI DR2 and SH$0$ES datasets, yielding $\beta = 0.256^{+0.0924}_{-0.0629}$ and $\beta = 0.335^{+0.0815}_{-0.0541}$, respectively, compared to other dataset combinations which exclude either of these two. Previous works have also shown constraints on non-zero interaction; our analysis yields a comparatively higher value of it, when DESI or S$H_0$ES data were included in the analysis. Moreover, while some interacting dark sector models in the literature have required the introduction of sign-switching mechanisms at specific redshifts to fully constrain the interaction, our approach offers a more economical solution. It fully constrains the interaction strength without invoking any exotic modifications. Additionally, we obtain an upper bound on the slope of the scalar field self-interaction potential through the parameter $\alpha$. Our AIC analysis further shows that the inclusion of SH$_0$ES and DESI DR2 BAO provides comparatively higher statistical preference in terms of positive and weak evidence in favor of the Chameleon model over the standard $\Lambda$CDM framework. These results suggest a mild preference for dark sector interaction in these particular two combinations of datasets, opening a promising avenue for new physics that could be further explored through small-scale structure observations by upcoming surveys, such as \textit{Euclid} and \textit{SPHEREx}. 

Using the best-fit cosmological parameters obtained from the MCMC analysis for various dataset combinations, we extract the scalar field dynamics, which exhibit damped oscillations around the evolving minimum of the effective potential. From this, we compute the evolution of the effective dark energy equation of state parameter, $w_{\rm eff}$, which crosses the phantom divide near $z \sim 0.5$. Even though the features like dynamical dark energy and phantom crossing are similar to the DESI observation, the assumption of the field to be at rest at the minima of the effective potential today, leads to an effective dark energy equation state value close to $-1$ at the present epoch. This does not agree with the DESI observations and as result it reflects in the weak statistical evidence in favor of the model relative to $\Lambda$CDM when Planck+DESI DR2 BAO+ Pantheon+ dataset combination is considered. It is also important to note that the current statistical evidence for phantom crossing from the DESI dataset remains moderate, future observations will offer the precision necessary to either confirm this feature or rule it out definitively.

This weak/positive statistical preference for a non-zero interaction strength $\beta$, derived from the analysis of linear cosmological scales, naturally motivates the extension of this investigation to smaller, non-linear scales. The best-fit values for $\beta$ obtained here imply that the variation in the effective gravitational constant in the transition regime can be as large as $\sim 1/3$. Owing to the increased number of independent modes at these scales, the statistical power is significantly enhanced, offering the potential for a more robust detection of $\beta$ or tighter constraints. A non-zero $\beta$ would also produce distinct observational signatures, such as a differential infall rate between dark matter and baryonic matter onto virialized halos \cite{Kesden_2006,Keselman_2009,Secco_2018}. Accurately quantifying these effects will require dedicated $N$-body simulations. Such simulations must be adapted to incorporate the Chameleon interaction and appropriately model the differing forces acting on dark matter and baryons. We plan to address these aspects in a forthcoming study.

\begin{acknowledgments}
We thank Simeon Bird, Shadab Alam, and Ethan O. Nadler for providing valuable comments on the manuscript. We also thank Prolay K. Chanda for the discussions in understanding the particle physics aspect of the model. SD and VG acknowledge the DST-SERB Government of India grant CRG/2019/006147 for supporting the project. AB acknowledges support from the Science and Engineering Research Board (SERB) India via the Startup Research Grant SRG/2023/000378. We acknowledge the use of NOVA, the high-performance computing cluster situated at the Indian Institute of Astrophysics, Bangalore, where all of the numerical calculations in this paper were performed. 
\end{acknowledgments}

\bibliography{main}{}
\bibliographystyle{aasjournalv7}



\end{document}